\documentclass[a4paper,twocolumn,english,superscriptaddress,showpacs,aps]{revtex4}
\usepackage{amsmath}
\usepackage{amssymb}
\usepackage{graphicx}
\usepackage{color}
\usepackage{hyperref}

\begin{document}
\title{A model for Rayleigh-B\'{e}nard magnetoconvection}
\author{Arnab Basak}
\author{Krishna Kumar}
\affiliation{Department of Physics, Indian Institute of Technology, Kharagpur-721 302, India}
{*e-mail: {\it kumar@phy.iitkgp.ernet.in}}
 
\date{\today}

\begin{abstract}
A model for three-dimensional Rayleigh-B\'{e}nard convection in low-Prandtl-number fluids near onset with rigid horizontal boundaries in the presence of a uniform vertical magnetic field is constructed and analyzed in detail. The kinetic energy $K$, the convective entropy $\Phi$ and the convective heat flux ($Nu-1$) show scaling behaviour with $\epsilon = r-1$ near onset of convection, where $r$ is the reduced Rayleigh number. The model is also used to investigate various magneto-convective structures close to the onset. Straight rolls, which appear at the primary instability, become unstable with increase in $r$ and bifurcate to three-dimensional structures. The straight rolls become periodically varying wavy rolls or quasiperiodically varying  structures in time with increase in  $r$ depending on the values of Prandtl number $Pr$. They become irregular in time, with increase in $r$. These standing wave solutions bifurcate first to periodic and then quasiperiodic traveling wave solutions, as $r$ is raised further. The variations of the critical Rayleigh number $Ra_{os}$ and the frequency $\omega_{os}$ at the onset of the secondary instability with $Pr$ are also studied for different values of Chandrasekhar's number $Q$. 
\end{abstract}
\pacs{47.35.Tv, 47.20.Bp, 47.20.Ky}
\maketitle

\section{\label{sec:Intro}Introduction}
The role of magnetic field on thermal convection in low-Prandtl-number fluids~\cite{nakagawa57,nakagawa59,chandra_1961,fauve_etal_1981,knobloch_etal_1981,proctor_weiss_1982,bc82,fauve_etal_1984,fauve_etal_prl_1984,kumar_etal_1986,meneguzzi87,cb89,bc89,glatzmaier_etal_1999,julien_etal_2000,cioni_etal_2000,ao_2001,bm_2002,houchens_2002,cattaneo_etal_2003,rucklidge_2006,dawes_2007,varshney_baig_2008,podvigina_2010,yanagisawa_etal_2011,pk_2012,brk_2014} has received considerable attention because of its importance in geophysical and astrophysical problems. The vertical magnetic field delays the onset of primary instability~\cite{nakagawa57,chandra_1961}, while the  horizontal magnetic field delays the onset of oscillatory (secondary) instability~\cite{fauve_etal_1981,fauve_etal_1984,meneguzzi87,pk_2012}. The latter also makes the straight rolls align along its direction. The external magnetic field also reduces the convective heat flux~\cite{proctor_weiss_1982,cb89,cioni_etal_2000,ao_2001,brk_2014}. The scaling behaviour of kinetic energy, convective entropy and heat flux near onset were studied recently with {\it stress-free} top and bottom plates~\cite{brk_2014}. They are not investigated for the magnetoconvection with rigid horizontal plates.  The secondary instability also leads to interesting three dimensional dissipative structures. It is assumed that the Nusselt number $Nu$ and the mean convective entropy $\Phi$ $=$ $\frac{1}{2}\int \theta^2 dV$ also scale like the mean kinetic energy $K$ $=$ $\frac{1}{2}\int v^2 dV$. Recent numerical simulations~\cite{brk_2014}, however, show that $Nu$ scales with $\epsilon = [Ra/Ra_c(Q)-1]$ as $\epsilon^{\alpha}$ with $\alpha < 1$ near the onset, where $Ra_c$ is the critical value of the  Rayleigh number at the primary convection. 

In this paper, we present a model of Rayleigh-B\'{e}nard convection in metallic fluids in the presence of uniform vertical magnetic field with rigid, thermally conducting and electrically insulating  horizontal plates. We use the model to investigate various properties of  magnetoconvection near onset. The kinetic energy $K$ is found to be proportional to $\epsilon$ near onset.  The convective entropy $\Phi$ and the convective heat flux ($Nu - 1)$,  scale with $\epsilon$ as  $\epsilon^{0.9}$ and $\epsilon^{0.91}$, respectively. The oscillatory convection appears at the secondary instability as $r$ is raised slowly. The distance from the criticality ($r_{os}-1$) for oscillatory instability scales with the Prandtl number $Pr$ as $Pr^{1.3}$.  We have observed the possibility of traveling as well as standing wave solutions above the onset of secondary instability. They were found to be periodic as well as quasiperiodic in time. A new traveling convective pattern consisting of alternating oblique wavy rolls is also observed. The time averaged root mean square of the velocity and temperature fields are also investigated. 

\section{\label{sec:Hydro}Hydromagnetic system}
We consider a thin horizontal layer of electrically conducting Boussinesq liquid of mean density $\rho_0$, thickness $d$, kinematic viscosity $\nu$, thermal diffusivity $\kappa$, magnetic diffusivity $\lambda$, and thermal  expansion coefficient $\alpha$ confined between two parallel plates, which is uniformly heated from below and uniformly cooled from the top. An adverse  temperature gradient $\beta$ is maintained across the fluid layer and a uniform magnetic field $B_0$ is applied anti-parallel to the acceleration due to gravity $\bm{g}$. We have chosen a Cartesian coordinate system  with origin in the middle of the fluid layer, the $xy$- plane coincident with the horizontal plane. The unit vector $\bm{e_3}$ is directed along the vertically upward direction, which is considered to be the positive direction of the $z$- axis. The fluid is initially at rest and allows conduction of heat flux along the vertical direction. As soon as $\beta$ is raised above a critical value $\beta_c$ keeping $B_0$ fixed, magnetoconvection sets in.  We ignore the magnetic Prandtl number $Pm =\nu/\lambda$ which is generally of the order of $10^{-5}$ or smaller for terrestrial fluids.  The magnetohydrodynamics is then governed by the following dimensionless equations:
\begin{eqnarray}
&\partial_t\bm{v} + (\bm{v\cdot\nabla})\bm{v} = -\nabla p + \nabla^2\bm{v} + Q\partial_z\bm{b} + Ra\theta\bm{e_3},
\label{eq:hmsys1}\\
&\nabla^2\bm{b} = -\partial_z\bm{v},
\label{eq:hmsys2}\\
&Pr[\partial_t\theta + (\bm{v\cdot\nabla})\theta] = \nabla^2\theta + v_3,
\label{eq:hmsys3}\\
&\bm{\nabla\cdot v}=\bm{\nabla\cdot b}=0,
\label{eq:hmsys4}
\end{eqnarray}
where $\bm{v}$ $\thinspace(x,y,z,t)$ $\equiv$ $(v_1,v_2,v_3)$ is the velocity field, $\bm{b}$ $\thinspace(x,y,z,t)$ $\equiv$ $(b_1,b_2,b_3)$ is the induced magnetic field due to convection, $\theta\thinspace(x,y,z,t)$ is the deviation in the temperature field from the steady conduction profile.  Lengths, time, temperature field and induced magnetic field  are measured in units of fluid depth $d$, viscous diffusion time $d^2/\nu$, $\nu\beta d/\kappa$  and $B_0\nu/\lambda$, respectively. The magnetoconvection is controlled by  the three dimensionless parameters:  (i) the Rayleigh number $Ra = \alpha \beta gd^4/ \nu \kappa$, a measure of the buoyancy force, (ii) the Prandtl number $Pr = \nu/ \kappa$, and (iii) Chandrasekhar's number $Q = B_0^2d^2/4\pi \rho_0\nu\lambda$, which is a measure of the imposed magnetic field. The  rigid, thermally conducting and electrically insulating  horizontal plates leads to following boundary conditions~\cite{varshney_baig_2008}:  
\begin{equation}
v_1 = v_2 = v_3 = \theta=0,\; b_1 = b_2 = \frac{\partial b_3}{\partial z} = 0 \label{bc} 
\end{equation}
at $z = \pm 1/2$. 

\section{\label{sec:model}The model}
We now construct a low-dimensional model to investigate the essential features of Rayleigh-B\'{e}nard magnetoconvection with realistic boundary conditions near onset. First we eliminate the pressure term from Eq.~\ref{eq:hmsys1} by taking curl once on both sides of Eq.~\ref{eq:hmsys1}. The projection of the resulting equation on vertical axis gives an equation for the vertical vorticity $\omega_3$, which is  given as: 
\begin{equation}
\partial_t\omega_3 + (\bm{v\cdot\nabla})\omega_3 - (\bm{\omega\cdot\nabla})v_3 = \nabla^2\omega_3 + Q\bm{e_3}\bm{\cdot}[\bm{\nabla\times}(\partial_z\bm{b})]
\end{equation}

Operating by curl twice on Eq.~\ref{eq:hmsys1}, using the equation of continuity, and then projecting on vertical axis leads to
\begin{eqnarray}
\partial_t\nabla^2 v_3 &+& \bm{e_3\cdot\nabla\times}[(\bm{\omega\cdot\nabla})\bm{v}-(\bm{v\cdot\nabla})\bm{\omega}] \nonumber\\
&=&\nabla^4 v_3 + Ra \nabla_H^2\theta - Q \bm{e_3}\bm{\cdot}[\bm{\nabla\times\nabla\times}(\partial_z\bm{b})]
\end{eqnarray}

We now follow the procedure used by Niederl\"{a}nder et al.~\cite{niederlander_etal_1991} for making a model with no-slip conditions in the context of ferro-fluids. All the fields are taken to be periodic in horizontal plane. Chandrasekhar's functions are used in the vertical direction for the expansion of vertical velocity field so that the no-slip conditions can be applied on velocity field at $z = \pm1/2$. As the magnetic field is slaved to the velocity field, it can be easily computed using Eq.~\ref{eq:hmsys2} once we know the velocity modes. The expansion for the vertical velocity $v_3$, the vertical vorticity $\omega_3$ $\equiv$ $(\bm{\nabla \times v})_3$ and the convective temperature $\theta$, compatible with the boundary conditions (Eq.~\ref{bc}) are:

\begin{eqnarray}
v_3\thinspace(x,y,z,t) &=& [W_{101}\cos{(kx)}+W_{\bar{1}01}\sin{(kx)} \nonumber\\
&+& W_{111}(t)\cos{(kx)}\cos{(ky)} \nonumber\\
&+& W_{\bar{1}11}(t)\sin{(kx)}\cos{(ky)} \nonumber\\
&+& W_{1\bar{1}1}(t)\cos{(kx)}\sin{(ky)} \nonumber\\
&+& W_{\bar{1}\bar{1}1}(t)\sin{(kx)}\sin{(ky)}]C_1(\lambda_1z),
\end{eqnarray}

\begin{eqnarray}
\omega_3\thinspace(x,y,z,t)&=& [Z_{011}(t)\cos{(ky)} \nonumber\\
&+& Z_{0\bar{1}1}(t)\sin{(ky)}]\cos(\pi z) \nonumber\\
&+& [Z_{112}(t)\cos{(kx)}\cos{(ky)} \nonumber\\
&+& Z_{\bar{1}12}(t)\sin{(kx)}\cos{(ky)} \nonumber\\
&+& Z_{1\bar{1}2}(t)\cos{(kx)}\sin{(ky)} \nonumber\\
&+& Z_{\bar{1}\bar{1}2}(t)\sin{(kx)}\sin{(ky)}]\sin{(2\pi z)},
\end{eqnarray}

\begin{eqnarray}
\theta\thinspace(x,y,z,t)&=& [T_{101}\cos{(kx)}+T_{\bar{1}01}\sin{(kx)} \nonumber\\
&+& T_{111}(t)\cos{(kx)}\cos{(ky)} \nonumber\\
&+& T_{\bar{1}11}(t)\sin{(kx)}\cos{(ky)} \nonumber\\
&+& T_{1\bar{1}1}(t)\cos{(kx)}\sin{(ky)} \nonumber\\
&+& T_{\bar{1}\bar{1}1}(t)\sin{(kx)}\sin{(ky)}]\cos{(\pi z)} \nonumber\\
&+& T_{002}(t)\sin{(2\pi z)},
\end{eqnarray}
where 

\begin{equation}
C_1\thinspace(\lambda_1z)= 
\frac{\cosh{\lambda_1z}}{\cosh{\lambda_1/2}}-\frac{\cos{\lambda_1z}}{\cos{\lambda_1/2}} 
\end{equation}
is the first order Chandrasekhar's function~\cite{chandra_1961,niederlander_etal_1991} with $\lambda_1\approx 4.73$.  The horizontal velocities are then computed using the relations:

\begin{eqnarray}
\nabla_H^2 v_1 &=& -\partial_{xz} v_3 - \partial_y \omega_3\\ 
\nabla_H^2 v_2 &=& -\partial_{yz} v_3 + \partial_x \omega_3.
\end{eqnarray}
Once all the three components of the velocity field are known, the horizontal vorticities can be computed easily. 

By projecting the hydromagnetic system of equations [Eqs.~\ref{eq:hmsys1}-\ref{eq:hmsys4}] on these modes, we  get a model for magnetoconvection with no-slip, thermally conducting and electrically insulating boundary conditions. The model consists of nineteen coupled ordinary differential equations. The shear flow, in general, can not be fully expressed in terms of the vertical velocity and the vertical vorticity ~\cite{cb89,hgf95}. For example, the dimensionless shear stresses
$\sigma_{13} = \partial_1 v_3 + \partial_3 v_1$ and $\sigma_{23} = \partial_2 v_3 + \partial_3 v_2$ are nonzero, even if a part of the horizontal velocities $v_1$ and $v_2$ depend only on the vertical (the third) coordinate. However, the terms dependent purely on the vertical coordinate cannot be generated from the vertical vorticity $\omega_3$.  The model presented here considers the possibility of the shear generated by the vertical vorticity but ignores the  generation shear due to the perturbations independent of $\omega_3$.

The model is integrated by the standard fourth order Runge-Kutta (RK4) method with a dimensionless time step of $10^{-3}$. We set  $k = k_c(Q)$, where the critical wave number $k_c(Q)$ is known from Chandrasekhar's linear theory~\cite{chandra_1961}. We first determine the critical Rayleigh number $Ra_c$ from the model and  compare them with the well known results of Chandrasekhar. Table~\ref{table:Ra_c} enlists the values of critical Rayleigh number $Ra_c(Q)$ for different values of $Q$ as obtained from the model with the values known from the linear theory~\cite{chandra_1961}.  For $Q = 0$, the value of $Ra_c$ is $1728$, which is exactly equal to the value obtained by Niederl\"{a}nder. This value is within $1.2\%$ of the value obtained by the linear theory~\cite{chandra_1961}. The error in determination of the critical Rayleigh number from the model is within $5\%$ for $Q = 50$. We restrict ourselves upto  $Q = 100$ for which the maximum error in $Ra_c$ is less than $6.6\%$.  

\begin{table}[ht]
  \begin{center}
\def~{\hphantom{0}}
  \begin{tabular}{c|c|c|c|c}
\hline
 $Q$ & $k_c (Q)$ & Linear theory  & Model  &  Error\\
  &  & $Ra_c(Q)$ & $Ra_c(Q)$ & \\ 
\hline\hline
$0$ & $3.13$ & $1707.8$ & $1728$ & $1.18\%$  \\
\hline
$10$ & $3.25$ & $1945.9$ & $1926$ & $1.02\%$ \\
\hline
$50$ & $3.68$ & $2802.1$ & $2664$ & $4.93\%$ \\
\hline
$100$ & $4.00$ & $3757.4$ & $3510$ & $6.58\%$ \\
\hline
\end{tabular}
\caption{A comparison of the critical Rayleigh numbers obtained from Chandrasekhar's linear theory and the model.}
\label{table:Ra_c}
 \end{center}
 \end{table}

\begin{figure}[h]
\begin{center}
\resizebox{0.5\textwidth}{!}{%
  \includegraphics{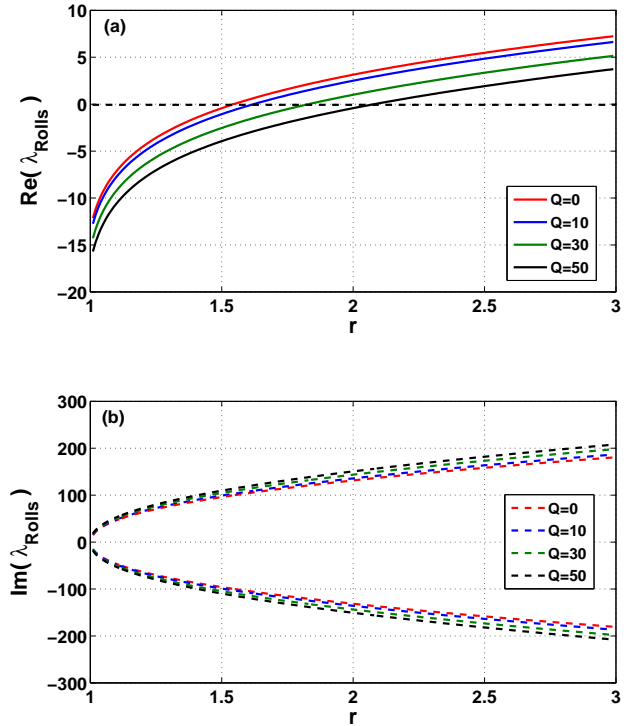}
}
\caption{\label{eigen_rolls_Q} (a) Real and (b) Imaginary parts of the largest eigenvalue as a function of $r$ for different values of $Q$ ($Pr = 0.1$). The onset of oscillatory instability is delayed and the frequency at the onset increases with increase in the Chandrasekhar's number $Q$.}
\end{center}
\end{figure}
 
\section{\label{sec:Stability}Stability of straight rolls}
For a thin layer of metallic fluid ($Pr > Pm$) confined between rigid, thermally conducting and electrically insulating horizontal boundaries, the magnetoconvection appear as  stationary straight rolls, as in the case of free-slip boundaries. The vertical magnetic field delays the onset of stationary convection. We now find the fixed points and their stability by analyzing the model. Just above the onset of straight rolls there are only five non-zero modes: $W_{101}$, $W_{\bar{1}01}$, $\theta_{101}$, $\theta_{\bar{1}01}$ and $\theta_{002}$. The roll fixed points are given by the following relations:
\begin{eqnarray}
W_{101}^* &=& f_1(r, Q, Pr)\theta_{101}^*, ~\ ~ W_{\bar{1}01}^* = f_1(r, Q, Pr) \theta_{\bar{1}01}^*, \nonumber\\
\theta_{002}^* &=& f_2 (r, Q, Pr), ~\ ~ \theta_{101}^{*2} + \theta_{\bar{1}01}^{*2} = \frac{c f_2}{f_1 Pr}, 
\end{eqnarray}
where 
\begin{eqnarray}
f_1 (r, Q, Pr) &=& \frac{a_1 r Ra_c(Q) k_c^2 (Q)}{a_2 + a_3 k_c^2 (Q) + a_4 k_c^4 (Q) + a_5 Q},\nonumber\\
f_2 (r, Q, Pr) &=& \frac{-b_1 + b_2 f_1 - b_3 k_c^2 (Q)}{b_4 f_1 Pr}  
\end{eqnarray}
with
$a_1$ $=$ $1.376 \times 10^{10}$, $a_2$ $=$ $9.881 \times 10^{12}$, $a_3$ $=$ $4.857 \times 10^{11}$, $a_4$ $=$ $1.974 \times 10^{10}$, $a_5$ $=$ $1.895 \times 10^{11}$, $b_1$ $=$ $9.741 \times 10^{10}$, $b_2$ $=$ $1.377 \times 10^{10}$, $b_3$ $=$ $9.869 \times 10^9$, $b_4$ $=$ $5.032 \times 10^{10}$, $c$ $=$ $15.485$. The temperature mode $T_{002}$ is the only nonlinear mode for the fixed point in the form of straight rolls. We now compare these modes with the expansion of the fields given by Clever and Busse~\cite{cb89} with no-slip horizontal boundaries. These five modes are exactly the same as the modes obtained by truncating the expansion of the velocity field at the first term and that of the convective temperature at the  second term.  Small errors in the  critical Rayleigh number $Ra_c$ for small values of $Q$ suggest that the model describes the qualitative features of magnetoconvection near the onset well.

\begin{figure}[h]
\begin{center}
\resizebox{0.5\textwidth}{!}{%
  \includegraphics{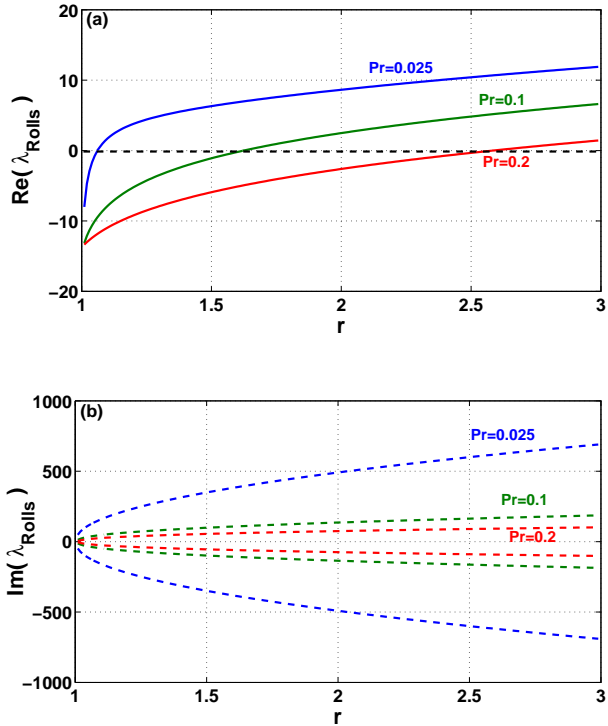}
}
\caption{\label{eigen_rolls_Pr} (a) Real and (b) Imaginary parts of the largest eigenvalue as a function of $r$ for different values of $Pr$ ($Q = 10$). The onset of oscillatory convection is delayed and the frequency at the onset decreases with increase in the Prandtl number $Pr$.}
\end{center}
\end{figure}

Rolls could be in any direction in an extended layer of metallic fluid. As the reduced Rayleigh number $r$ is raised slowly in steps, the perturbations in the form of vertical vorticity may be excited through nonlinear interaction with the vertical velocity. We investigate the stability of rolls in the presence of additional fourteen modes. We find the eigenvalues of a $19 \times 19$ matrix, obtained by linearizing about the roll fixed points. The eigenvalue $\lambda_m$ with the largest real part is found to form a complex conjugate pair.  

Figure~\ref{eigen_rolls_Q} shows the variation of (a) the real  and (b)  imaginary parts of $\lambda_m$ with $r$ for different values of $Q$ for $Pr = 0.1$. The real part of $\lambda_m$ becomes positive at greater value of $r$ for larger $Q$ values. This suggests that the oscillatory instability via forward Hopf bifurcation is also delayed by the vertical magnetic field. The frequency $\omega_{os}$ at the onset of oscillatory instability shifts to higher values, as $Q$ increases. 

\begin{figure}[h]
\begin{center}
\resizebox{0.5\textwidth}{!}{%
  \includegraphics{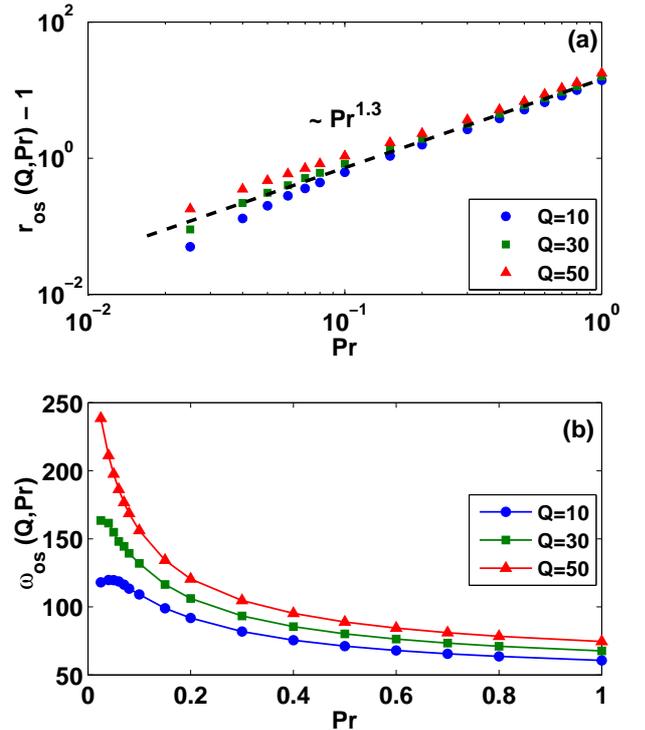}
}
\caption{\label{scaling_Pr_Q} Variation of (a) the critical value of the reduced Rayleigh number $r_{os} (Q, Pr)-1$ and (b) frequency at the onset of oscillatory magnetoconvection as a function of  as a function of the Prandtl number $Pr$ for different values of $Q$.}
\end{center}
\end{figure}

The onset of oscillatory instability is greatly affected by the variation of the Prandtl number $Pr$. Figure~\ref{eigen_rolls_Pr} displays the variation of (a) the real  and (b)  imaginary parts of $\lambda_m$ with $r$ for different values of $Pr$ for $Q = 10$. The onset of oscillatory instability shifts to a larger value with $Pr$ and the frequency at the onset decreases with increasing values of $Pr$, which is consistent with the behaviour in the absence of any magnetic field. Figure~\ref{scaling_Pr_Q} (a) displays the variation of the threshold for oscillatory instability $\epsilon_{os} = r_{os} (Q, Pr)-1$ with 
Prandtl number $Pr$ for different values of $Q$. The threshold $r_{os}$ scales with $Pr$ as $Pr^{1.3}$ for $Pr \geq 0.1$ and the scaling appears to be independent of $Q$. The value of $r_{os}$ is larger for higher values of $Q$ at a fixed value of $Pr$. The frequency $\omega_{os}$ at the onset of oscillatory instability increases with increase in $Q$ at a fixed value of $Pr$. The frequency $\omega_{os}$ decreases with $Pr$ for $Pr \ge 0.1$. However, the variation of the frequency with $Pr$ is non-monotonic for much smaller values ($Pr < 0.1$). 

\section{\label{sec:Scaling}Scaling of global quantities near onset}

We then use the model to investigate possible scaling behaviour near the onset of magnetoconvection.  We start integration of the hydromagnetic system with randomly chosen initial conditions for a given value of $Pr$ and $r$. The value of $r$ is raised in small steps keeping $Pr$ fixed. The final values of all the fields of the last run are then used as initial conditions of a fresh run. As $Pr$ is always greater than $Pm$, which is assumed to be vanishingly small, we always observe stationary straight (two-dimensional) rolls at the primary instability. This is consistent with the Chandrasekhar's prediction. We find various time dependent dissipative structures at secondary and higher order instabilities, as $r$ is raised further. The whole process is repeated for different values of $Pr$. We have also checked several integration with random initial conditions to find out any possibility of hysteresis. We did not find any hysteresis in the model.


\begin{figure}[h]
\begin{center}
\resizebox{0.5\textwidth}{!}{%
  \includegraphics{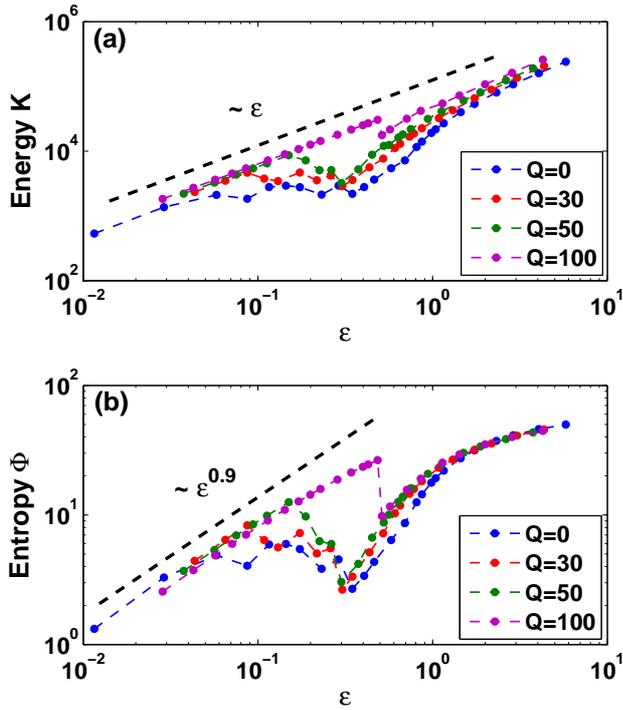}
}
\caption{\label{fig:Ek_En_P1} Variations of (a) the mean kinetic energy per unit mass $K$ and (b) the mean convective entropy $\Phi$ with the parameter $\epsilon = [Ra/Ra_c(Q)-1]$ for $Pr = 0.025$, as obtained from the model.}
\end{center}
\end{figure}


We now present the results of scaling behaviour of three global quantities: the kinetic energy per unit mass $K = \frac{1}{2}\int v^2 dV$,  the convective entropy per unit mass $\Phi$ $=$ $\frac{1}{2}\int \theta^2 dV$, and the convective heat flux ($Nu-1$) with $\epsilon = [Ra/Ra_c(Q)-1]$ for different values of Chandrasekhar's number $Q$. The parameter $\epsilon$ is a measure of the distance from the criticality. Figure~\ref{fig:Ek_En_P1}(a) shows the the kinetic energy $K$  as a function  $\epsilon$ for $Pr = 0.025$ and for four different values of $Q$.   The kinetic energy increases linearly with $\epsilon$ for time independent magnetoconvection.  This means that the average speed of the fluid flow is proportional to $\sqrt{\epsilon}$ at the primary instability, which is a well known result. The value of $K$ is higher for larger values of $Q$. The kinetic energy shows a sharp decrease, as $\epsilon$ is raised in small steps.  The sharp decrease in the kinetic energy is accompanied by  a time dependent magnetoconvection. The convection is found to be quasiperiodic in time at the onset of secondary instability. The time-averaged value of the kinetic energy first decreases, attains a minimum, and then  increases once again  with  increase in $\epsilon$. The sharp decrease in $K$ occurs at higher values of $\epsilon$ for larger values of $Q$. The dip in $K$ is also shallower at higher values of $Q$. The larger vertical magnetic field delays the onset of secondary instability, which is time-dependent. As $\epsilon$ is raised further, time averaged value of $K$ is again found to be varying almost linearly with $\epsilon$ for $1 \le \epsilon < 10$. 

Figure~\ref{fig:Ek_En_P1}(b) shows the variation of the convective entropy $\Phi$ with $\epsilon$ for $Pr=0.025$. The qualitative behaviour is similar to that of the kinetic energy but with some significant differences: $\Phi$ scales with $\epsilon$ as $\epsilon^{0.9}$ near the onset. The convective temperature field is then proportional to $\epsilon^{0.45}$ which is different than the behaviour of the average speed. This is due to the fact that a part of the available thermal energy is used to maintain a net thermal flux in the vertically upward direction. There is no net momentum flux in the vertical direction. The scaling behaviours of the average kinetic energy $K$ is slightly different from that of the convective entropy $\Phi$. The entropy also shows a sharp decrease at the onset of time-dependent (secondary) instability. For higher values of $\epsilon$ ($1 < \epsilon < 10$),  all the curves for $\Phi (\epsilon)$ have a common slope approximately equal to $0.4$, which is much smaller than its value ($\approx 0.9$) just above the onset of magnetoconvection. 

\begin{figure}[h]
\begin{center}
\resizebox{0.5\textwidth}{!}{%
  \includegraphics{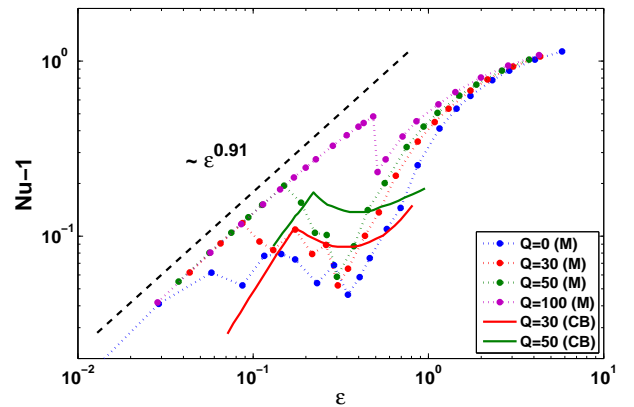}
}
\caption{\label{fig:Nu_ep_CB_P1} Plot of convective heat flux ($Nu-1$) versus $\epsilon = [Ra-Ra_c(Q)]/Ra_c(Q)$ for $Pr = 0.025$ as obtained from the model (M) and compared with the results of Clever and Busse (CB)(1989).}
\end{center}
\end{figure}

The quantity $Nu-1$ is a measure of the convective heat flux across the fluid layer, where the Nusselt number is defined as: $Nu$ $=$ $1 + Pr^2 <v_3\theta>_{xyz}$. The symbol $<>_{xyz}$ stands for the spatial average. Figure~\ref{fig:Nu_ep_CB_P1} shows the scaling behaviour of the time averaged value of the convective heat flux with $\epsilon$ for $Pr = 0.025$. The points show the results obtained from the model for different values of $Q$ and the dashed line is parallel to the best fit over data points near the onset. The best fit shows that the convective heat flux scales with $\epsilon$ as $\epsilon^{0.91}$ near the primary instability.  There is a sharp fall in the heat flux at the onset of time-dependent convection. The time averaged value of the convective heat flux then starts increasing initially much faster and then much slowly with increase in $\epsilon$. The variation  of the convective heat flux with $\epsilon$ becomes almost identical for all values of $Q$ investigated at higher values of $\epsilon$. In this regime, the slope the $Nu - \epsilon$ curve at higher values of $\epsilon$ is lower than $0.91$. The continuous curves show the variation of convective heat flux with $\epsilon$, as reported by Clever and Busse~\cite{cb89} for $Pr = 0.025$ with no-slip boundary conditions. They found the scaling exponent to be much larger ($ > 1.4$) for the stationary magnetoconvection, which is unusual. The qualitative behaviour of the results obtained from the model has broad similarity with those obtained by Clever and Busse~\cite{cb89}, but our model always shows the scaling exponent of heat flux with $\epsilon$ less than unity near the onset of stationary convection. It is also in excellent agreement with the recent results from direct numerical simulations with free-slip boundary conditions~\cite{brk_2014}. No experiment suggests that the  scaling behaviour $Nu-1 \sim \epsilon^{\alpha}$ with the scaling exponent $\alpha \ge 1.4$ near the primary instability (stationary convection). Like other global variables, the heat flux also decreases at the onset of oscillatory instability, reaches a minimum and then increases with $\epsilon$. The scaling exponent is less than unity for $\epsilon > 1 (r > 2)$. 

\begin{figure}[h]
\begin{center}
\resizebox{0.5\textwidth}{!}{%
  \includegraphics{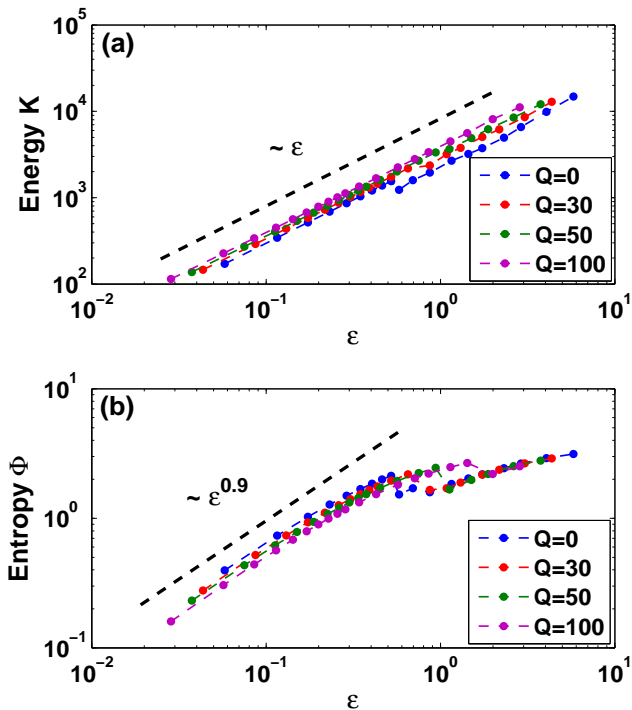}
}
\caption{\label{fig:Ek_En_P2} Variations of (a) the mean kinetic energy $K$ and (b) the mean convective entropy $\Phi$ with $\epsilon$ for $Pr = 0.1$ as obtained from the model.}
\end{center}
\end{figure}

\begin{figure}[h]
\begin{center}
\resizebox{0.5\textwidth}{!}{%
  \includegraphics{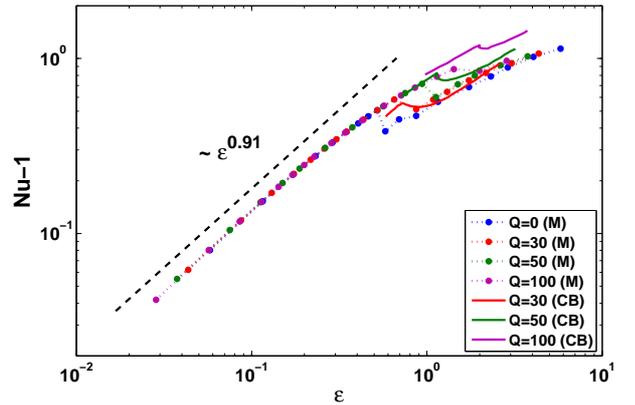}
}
\caption{\label{fig:Nu_ep_CB_P2} Plot of convective heat flux ($Nu-1$) versus $\epsilon = [Ra-Ra_c(Q)]/Ra_c(Q)$ for $Pr = 0.1$ as obtained from the model (M) and obtained by Clever and Busse (CB)(1989). }
\end{center}
\end{figure}

\begin{table*}[ht]
  \begin{center}
\def~{\hphantom{0}}
  \begin{tabular}{c|c|c|c|c|c|c|c|c}
\hline
Fluid  &\multicolumn{4}{c|}{$Pr=0.025$}&\multicolumn{4}{c}{$Pr=0.1$} \\
\cline {2-5} \cline{6-9}
patterns  & r (Q = 0) & r (Q = 10) & r (Q = 30) & r (Q = 50) & r (Q = 0) & r (Q = 10) & r (Q = 30) & r (Q = 50)\\ 
\hline\hline
2D Rolls & $\leq 1.04$ & $\leq 1.05$ & $\leq 1.09$ & $\leq 1.18$ & $\leq 1.54$ & $\leq 1.62$ & $\leq 1.82$ & $\leq 2.07$ \\
\hline
WR & $-$ & $-$ & $-$ & $-$ & $1.55 - 1.62$ & $1.63$ & $-$ & $-$ \\
\hline
QWR & $1.05 - 1.11$ & $1.06 - 1.12$ & $1.10 - 1.14$ & $-$ & $1.63 - 1.64$ & $1.64$ & $-$ & $-$ \\
\hline
CWR & $1.12 - 1.14$ & $1.13 - 1.27$ & $1.15 - 1.26$ & $1.19 - 1.26$ & $1.65 - 1.75$ & $1.65 - 1.77$ & $1.83 - 1.85$ & $-$ \\
\hline
QWR & $1.15 - 1.18$ & $-$ & $-$ & $-$ & $-$ & $-$ & $-$ & $-$ \\
\hline
CWR & $1.19 - 1.29$ & $-$ & $-$ & $-$ & $-$ & $-$ & $-$ & $-$ \\
\hline
TW	& $1.30 - 1.31$ & $1.28 - 1.30$ & $1.27 - 1.29$ & $1.27 - 1.29$ & $1.76 - 1.77$ & $1.78 - 1.79$ & $1.86 - 1.88$ & $-$ \\
\hline
QTW & $-$ & $-$ & $1.30$ & $1.30$ & $1.78 - 1.92$ & $1.80 - 1.95$ & $1.89 - 2.04$ & $2.08 - 2.13$ \\
\hline
CTW	& $\geq 1.32$ & $\geq 1.31$ & $\geq 1.31$ & $\geq 1.31$ & $\geq 1.93$ & $\geq 1.96$ & $\geq 2.05$ & $\geq 2.14$ \\
\hline
 \end{tabular}
\caption{Magnetoconvective patterns obtained from the model for $Pr = 0.025$ and $Pr = 0.1$. Two-dimensional stationary rolls (2D Rolls), periodic wavy rolls (WR), quasiperiodic wavy rolls (QWR), chaotic wavy rolls (CWR), periodic traveling waves (TW), quasiperiodic traveling waves (QTW) and chaotic traveling waves (CTW) are observed in the model.}
\label{flow-patterns}
 \end{center}
 \end{table*}

Figure~\ref{fig:Ek_En_P2}(a) shows the variation of mean kinetic energy $K$  as a function  $\epsilon$ for $Pr = 0.1$ and for different values of $Q$. The kinetic energy $K$ varies again linearly with $\epsilon$ near the onset of the stationary convection in the form of straight rolls, and decreases at the onset of time-periodic instability. However, oscillatory instability occurs at much higher values or $r$ in this case, and the decrease in $K$ at the onset of oscillatory instability is small. Immediately after the secondary instability, $K$ increases linearly with $\epsilon$. Figure~\ref{fig:Ek_En_P2}(b) displays the variation of $\Phi$ with $\epsilon$ for $Pr = 0.1$.  The global variable $\Phi$ scales with $\epsilon$ as $\epsilon^{0.9}$ near the primary instability. All the curves for $\Phi (\epsilon)$ again merge for $1 < \epsilon < 10$. The scaling exponent of $\Phi$ in this region is found to be $0.4$.

Figure~\ref{fig:Nu_ep_CB_P2} shows the plot of $Nu-1$ versus $\epsilon$ for $Pr = 0.1$. The points connected by dots are the results obtained from the model for  different values of Chandrasekhar's number ($Q = 0,30,50$ and $100$). Three solid lines are the results from DNS (Clever and Busse)~\cite{cb89} with no-slip conditions for $Q = 30$, $50$ and $100$. The linear dashed line is the best fit of data points  obtained from the model for magnetoconvection near the onset. The convective heat flux shows the same scaling behaviour as before in the regime of stationary magnetoconvection. The results obtained from the model are in better agreement with those of Clever and Busse~\cite{cb89} for $Pr = 0.1$. For higher values of $\epsilon$, all the curves seem to become parallel to each other. This is in agreement with the results of DNS.

\section{\label{sec:Patterns}Time-dependent fluid patterns}

As the Rayleigh number is increased,  different time dependent patterns appear at the secondary instability for different values of $Pr$. The threshold for the secondary instability increases as $Q$ is raised.  For low-Prandtl-number convection~\cite{meneguzzi87,cb81}, the secondary instability is oscillatory convection, which occurs close to the onset of primary convection. This suggests that a qualitative description of  the convection near secondary instability in the presence of small magnetic field may be captured qualitatively with relatively less number of modes.  The selection of the modes for the vertical vorticity are done to construct a minimum mode model to be able to capture the scaling behaviour of the global quantities of Rayleigh-B\'{e}nard magnetoconvection near the onset.  The scaling properties are captured quite well in the model.  The fluid patterns are however sensitive to $Pr$, $r$, $Q$ and wavenumber of the perturbations. A single model
is unlikely to capture the patterns with variations of all these parameters. In addition, we have considered only the critical wave number $k_c (Q)$. We have tested the model by adding more vorticity modes. The  oscillatory nature of the secondary stability does not change qualitatively. However, the onset of higher order (e.g. tartiary) instability is affected. We now discuss the magnetoconvective patterns obtained from the model presented here. 

We observe temporal quasiperiodic wavy rolls (QWR) for  $Pr = 0.025$ and periodic wavy rolls (WR) for $Pr = 0.1$ at the onset of secondary instability. In the latter case, the convection becomes temporally quasiperiodic with increase in $r$. Further increase in $r$ leads to chaotic wavy rolls (CWR).  The onset of chaotic waves also delayed for larger value of $Q$. For a fixed value of $Q$ the onset of secondary instability is higher for higher value of $Pr$. At relatively higher values $r$, the time dependent convective flow bifurcates from standing waves to traveling waves. The convective flow consists of periodic traveling waves (TW) in a narrow range of $r$, which depends on $Pr$ and $Q$. Wavy rolls travel in a direction perpendicular to the direction of roll-axis. Further increase in $r$ leads to quasiperiodic traveling waves (QTW) and chaotic traveling waves (CTW). Table~\ref{flow-patterns} lists the fluid patterns observed for different values of $r$, $Pr$ and $Q$.

\begin{figure}[h]
\begin{center}
\resizebox{0.5\textwidth}{!}{%
  \includegraphics{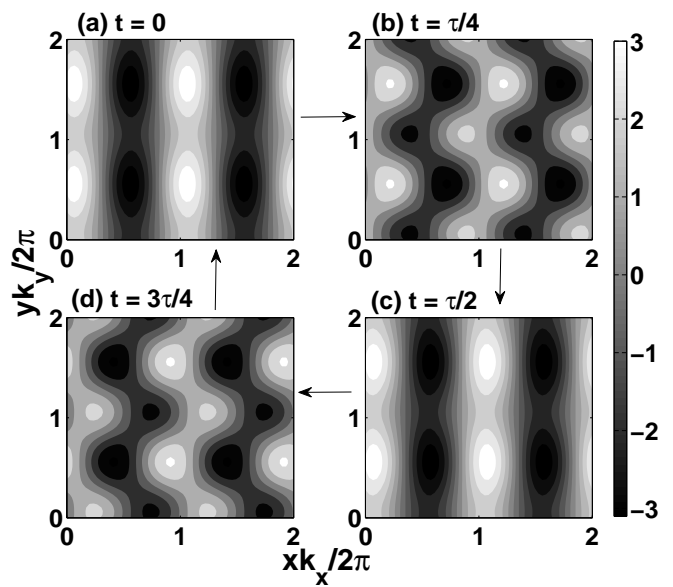}
}
\caption{\label{fig:wavy_contour} Contour plots of the convective temperature field at mid-plane $z=0$ showing wavy rolls (WR) for $Pr=0.1$, $Q=10$ ($k_c=3.25$) and $r=1.63$ at four instants (a) $t=0$, (b) $t=\tau/4$, (c) $t=\tau/2$, and (d) $t=3\tau/4$, where $\tau=0.88$ is the dimensionless time period of the wavy rolls.}
\end{center}
\end{figure}

\begin{figure}[h]
\begin{center}
\resizebox{0.5\textwidth}{!}{%
  \includegraphics{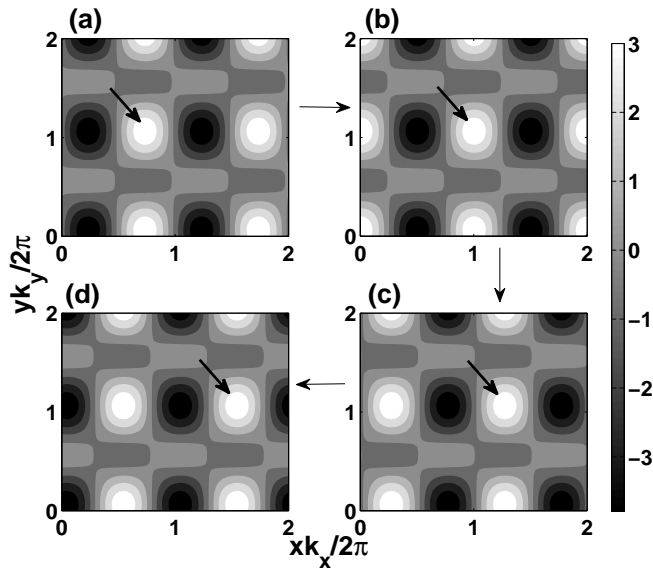}
}
\caption{\label{fig:tw_contour} Contour plots of the convective temperature field at the mid-plane $z=0$ showing periodic traveling waves (TW)  for $Pr=0.1$, $Q=10$ ($k_c = 3.25$) and $r = 1.78$ at (a) $t=0$, (b) $t=0.40$, (c) $t=0.80$, and (d) $t=1.20$. The inclined arrows show the positions of the same hot (white) region with time. The fluid patterns move towards the right.}
\end{center}
\end{figure}

We now discuss the results on the fluid patterns observed for $Pr=0.1$ and $Q=10$. Figure~\ref{fig:wavy_contour} shows the contour plots of the convective temperature field at mid-plane $z=0$ for $r=1.63$ at four different instants. The convection shows wavy rolls (WR). They represent standing wave solutions. The positions of nodes do not vary in time. The dimensionless period oscillation for this periodic standing wave solution is $\tau=0.88$. Figure~\ref{fig:tw_contour} shows the mid-plane contour plots of the convective temperature field for $r=1.78$ at four instants (a) $t=0$, (b) $t=0.40$, (c) $t=0.80$, and (d) $t=1.20$.  Convective patterns are shown to travel slowly along the positive $x-$ direction. The inclined arrows mark the position of the same hot (white) region which now moves with time. The structures along the $y$-axis make the patterns three dimensional, which travel from left to right in this viewgraph. 

\begin{figure}[h]
\begin{center}
\resizebox{0.5\textwidth}{!}{%
  \includegraphics{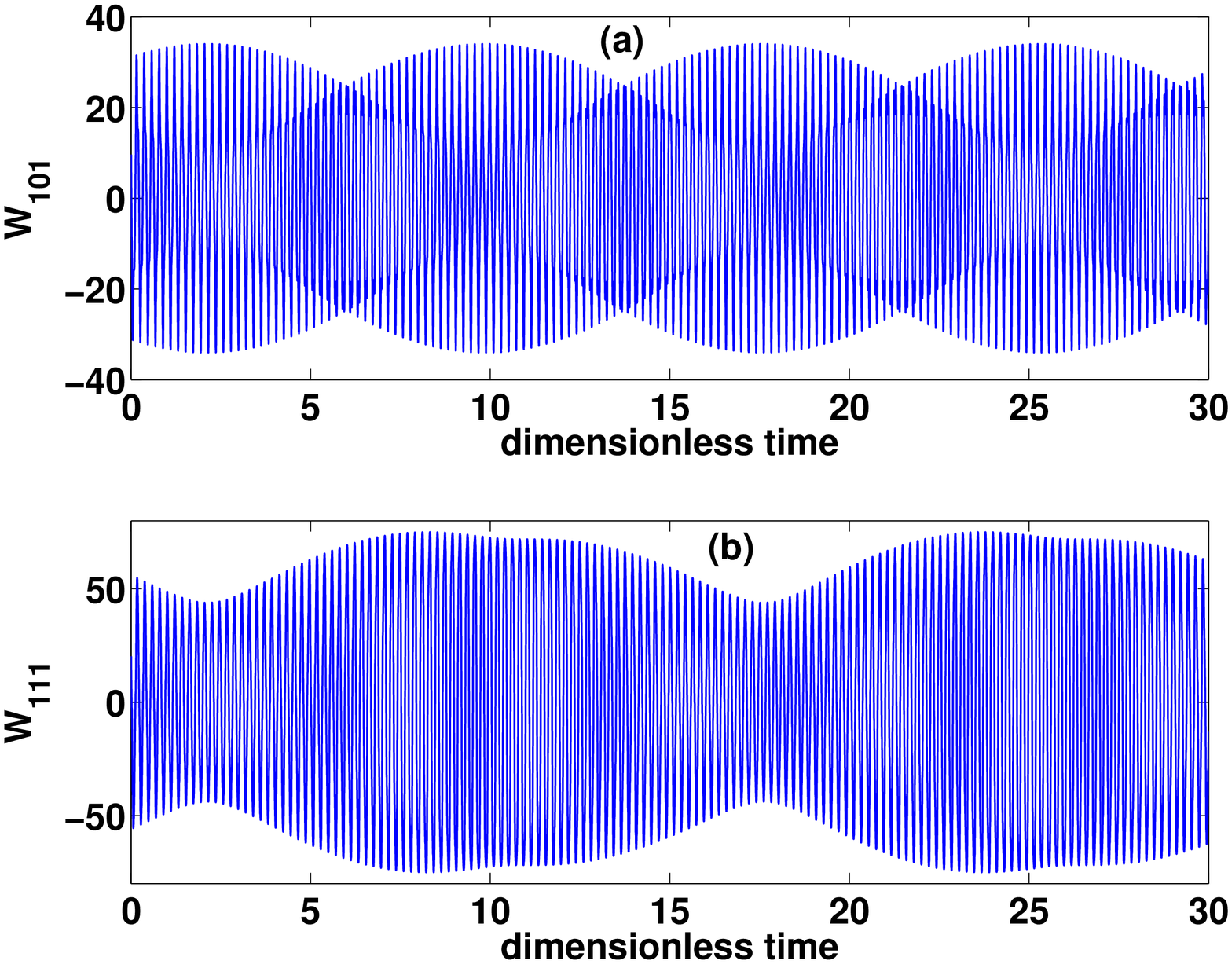}
}
\caption{\label{fig:qtw_signal} The temporal variation of the two largest Fourier modes (a) $W_{101}$ and (b) $W_{111}$  for $Pr =0.1$, $Q=10$ and $r=1.90$ [$k_c (Q)= 3.25$] showing a quasiperiodic.}
\end{center}
\end{figure}

\begin{figure}[h]
\begin{center}
\resizebox{0.5\textwidth}{!}{%
  \includegraphics{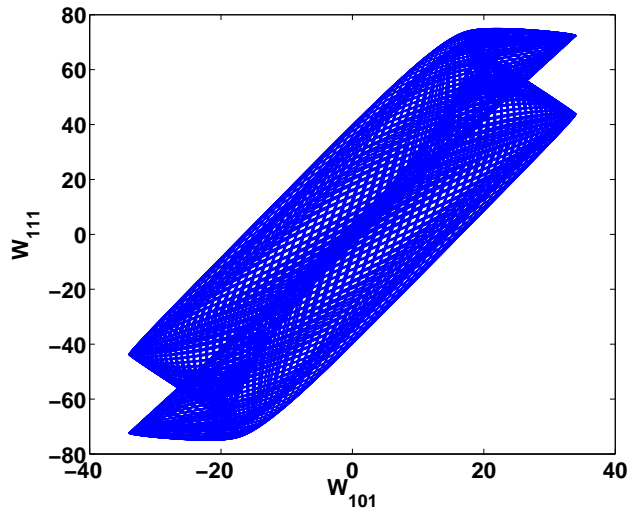}
}
\caption{\label{fig:phase_portrait} The projection of the phase space on the $W_{111}-W_{101}$ plane  showing  quasiperiodic magnetoconvection. All parameters are same as given in Fig.~\ref{fig:qtw_signal}. }
\end{center}
\end{figure}

\begin{figure}[h]
\begin{center}
\resizebox{0.5\textwidth}{!}{%
  \includegraphics{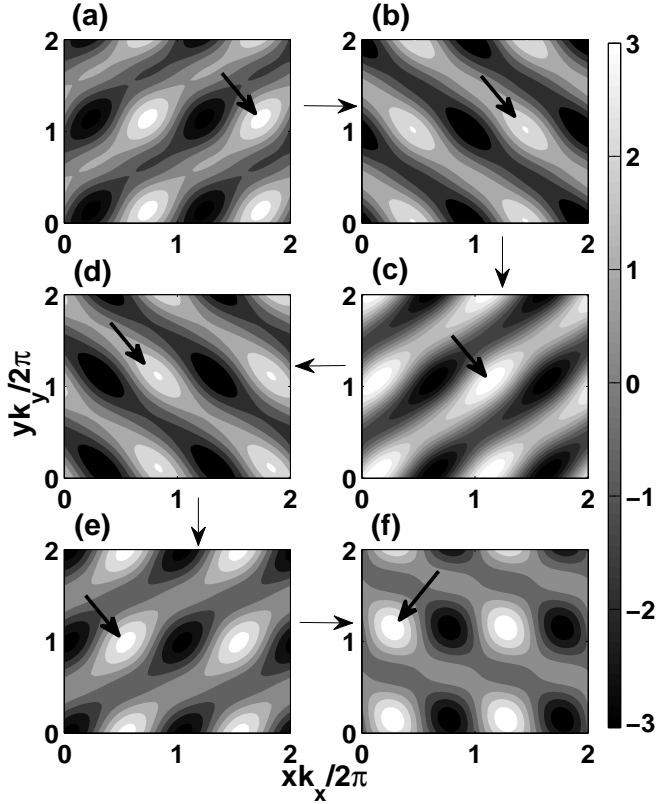}
}
\caption{\label{fig:qtw_contour} Contour plots of the convective temperature field at mid-plane $z=0$ for $r=1.9$ showing quasiperiodic traveling waves (QTW) at several instants: (a) $t=0$, (b) $t=0.60$, (c) $t=1.20$, (d) $t=1.80$, (e) $t=2.40$, and (f) $t=3.00$. The inclined arrows show the positions of the same hot region with time. Other parameters are as given in Fig.~\ref{fig:qtw_signal}.}
\end{center}
\end{figure}

As $r$ is raised further, the periodic traveling waves become quasiperiodic in time. Figure~\ref{fig:qtw_signal} shows the temporal variation of the two largest Fourier modes: (a) $W_{101}$ and (b)$W_{111}$ for $r = 1.9$, $Pr =0.1$ and $Q=10$. These signals suggest quasiperiodic magnetoconvection. The period of amplitude modulation for the Fourier mode $W_{111}$ is almost double of the time of amplitude modulation of the Fourier mode $W_{101}$. Figure~\ref{fig:phase_portrait} shows the projection of the phase space of the hydro-magnetic system on the $W_{111}-W_{101}$ plane. As time passes, a particular region of the phase space gets filled completely. This confirms the convection to be quasiperiodic. 

Figure~\ref{fig:qtw_contour} displays the mid-plane contour plots of the convective temperature field for this case ($r=1.9$) at  different instants of time. The patterns consists of inclined wavy rolls. The axis of the wavy rolls keep alternating in time between mutually perpendicular directions. The system of alternating oblique wavy rolls also keep traveling along the negative direction of the $x$-axis. The inclined arrows shown in the figure follow the positions of the same hot region, which moves towards the left as time passes. This complex and new magnetoconvective patterns are quasiperiodic in time. 
We have not investigated the possibility of convection in the form of square patterns. They may be possible at relatively higher values of $r$. In the DNS with stress-free boundaries~\cite{brk_2014}, two sets of stationary rolls leading to patterns of asymmetric squares (cross-rolls) appear at relatively higher values $r$ at the tartiary instability. 

\begin{figure}[h]
\begin{center}
\resizebox{0.5\textwidth}{!}{%
  \includegraphics{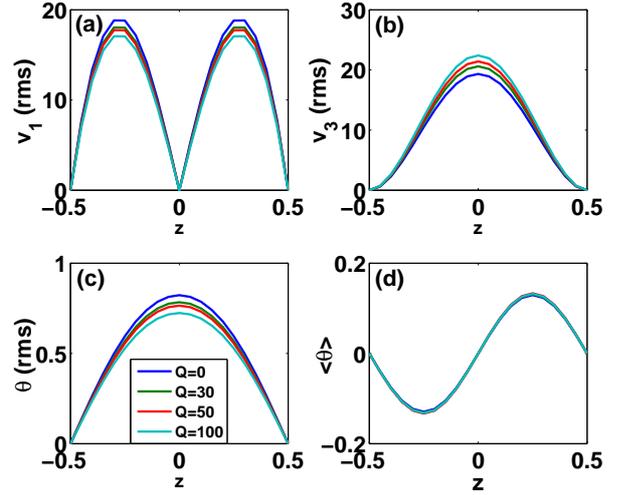}
}
\caption{\label{rms_variation_Q} Variations of (a) $v_1\thinspace(rms)$ and (b) $v_3\thinspace(rms)$, (c) $\theta\thinspace(rms)$, and (d) $<\theta>$ along $z-$ axis for stationary 2D convection ($Pr=0.1$, $r=1.05$) for different values of $Q$. The velocity $v_2\thinspace(rms)$ is always zero for 2D rolls parallel to the $y$-axis.}
\end{center}
\end{figure}

\begin{figure}[h]
\begin{center}
\resizebox{0.5\textwidth}{!}{%
  \includegraphics{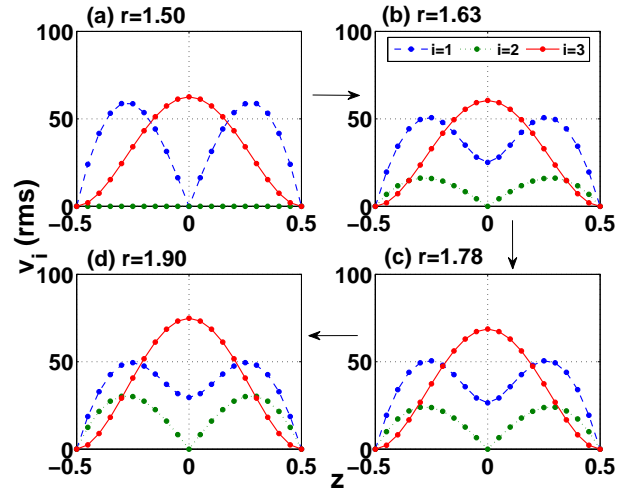}
}
\caption{\label{rms_variation_r} Variations of root mean square velocities in the vertical ($z$) direction for four different values of $r$: (a) $r=1.50$, (b) $r=1.63$, (c) $r=1.78$, (d) $r=1.90$. $v_1\thinspace(rms) (z)$, $v_2\thinspace(rms) (z)$ and $v_3\thinspace(rms) (z)$ are  represented by blue dashed-dotted curve, green line with dots and  red curve with dots, respectively. Other parameters are: $Pr=0.1$, $Q=10$ ($k_c=3.25$).}
\end{center}
\end{figure}
\begin{figure}[h]
\begin{center}
\resizebox{0.5\textwidth}{!}{%
  \includegraphics{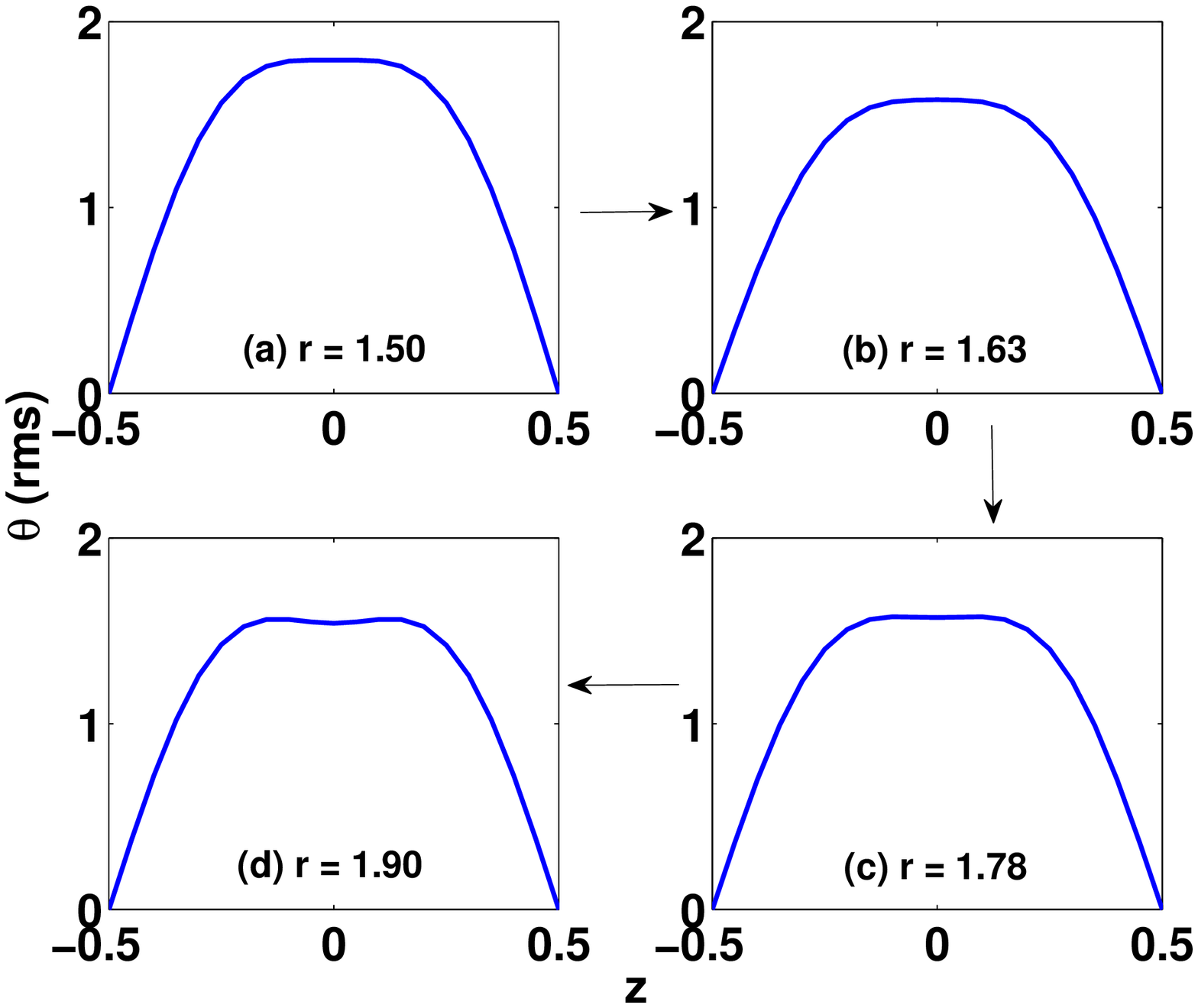}
}
\caption{\label{temp_variation_r} Variation of root mean square value of the convective temperature field in the vertical direction for four different values of $r$: (a) $r=1.50$, (b) $r=1.63$, (c) $r=1.78$, (d) $r=1.90$. Other parameters are: $Pr = 0.1$, $Q = 10$, $k_c (Q) = 3.25$.}
\end{center}
\end{figure}

\section{\label{subsec:rms_quantities}Convective temperature and velocity profiles}

We now present the results on the temperature and velocity profiles obtained from the model. We define a symbol $<...>_{x,y,t}$ to describe the spatial average in horizontal plane and the temporal average of any quantity inside the angular bracket. We then study the root mean square (rms) of any relevant quantity $f$ as $f\thinspace(rms) = \sqrt{<f^2>_{x,y,t}}$, which is a function of the vertical coordinate $z$. Figure~\ref{rms_variation_Q} shows the variations of (a) $v_1\thinspace(rms)$ and (b) $v_3\thinspace(rms)$ along the vertical axis for stationary straight rolls in fluid with $Pr=0.1$ at $r=1.05$ for different values of $Q$. $v_2\thinspace(rms)$ is always zero for straight rolls parallel to $y$-axis. The peak of the horizontal velocity field [$v_1\thinspace(rms)$] decreases and that for the vertical velocity field  [$v_3\thinspace(rms)$] increases with increase in $Q$. The variation in the rms value of the convective temperature $\theta\thinspace(rms)$ in  the vertical direction for straight rolls is shown in Fig.~\ref{rms_variation_Q} (c). The rms of the convective temperature has a peak  in the middle of the cell just above the onset of convection, as in the case of the vertical velocity. However, the peak rms value decreases with increase in $Q$ value. The temporal mean of horizontally averaged convective temperature field is more or less sinusoidal [Fig.~\ref{rms_variation_Q} (d)]. The effect of the vertical magnetic field is negligible for the $Q$ values investigated here. 

Figure~\ref{rms_variation_r} shows the variations of $v_1\thinspace(rms)$ (blue dashed line with points), $v_2\thinspace(rms)$ (green dotted line with points), and $v_3\thinspace(rms)$ (red solid line with points) along the vertical direction for $Pr=0.1$, $Q=10$ ($k_c (Q) = 3.25$) and for different values of values of $r$. Figure~\ref{rms_variation_r}(a) describes the  case of stationary straight (2D) rolls along the $y$-axis for $r=1.50$, where $v_2\thinspace(rms)$ is always zero. $v_1\thinspace(rms)$ has two peaks: one in the upper part and another in the lower part of the cell. It vanishes in the middle of the cell ($z=0$).  The rms value of the vertical velocity $v_3\thinspace(rms)$ has a peak at $z=0$. All quantities are symmetric about $z=0$ and vanish at the top and bottom boundaries $z=\pm 0.5$ due to no-slip conditions. Fig.~\ref{rms_variation_r}(b) describes three-dimensional convection for $r=1.63$.   The velocity along the roll-axis $v_2\thinspace(rms)$ becomes non-zero, as three-dimensional convection begins. It shows two peaks located at the positions of peaks in $v_1\thinspace(rms)$, but with less in magnitude. The minima of $v_1\thinspace(rms)$ at $z=0$ is now non-zero while the minima of $v_2\thinspace(rms)$ is zero in the middle of the cell. The peak of the vertical velocity $v_3\thinspace(rms)$ at $z=0$ is slightly reduced.  With further increase in $r$,  the peak values of $v_2\thinspace(rms)$ and $v_3\thinspace(rms)$ [Figs.~\ref{rms_variation_r}(c) and (d)] also increase, while that of $v_1\thinspace(rms)$ remains almost constant. 

Figure~\ref{temp_variation_r} shows the variation of the rms value of the convective temperature in vertical direction as $r$ is varied keeping all other parameters fixed ($Q=10$ and $Pr=0.1$) for convective structures. The rms of the convective temperature, which had a peak in the middle at the onset of stationary convection, becomes flat with increase in $r$. Figure~\ref{temp_variation_r} (a) shows the rms of the convective temperature at $r=1.50$. The rms value 
value of the convective temperature drops in the middle of the cell at the onset of oscillatory convection [Fig.~\ref{temp_variation_r} (b)], which is expected. As $r$ is raised further, the convective temperature shows bimodal
behavior [Figs.~\ref{temp_variation_r} (c) and (d)].

\section{\label{sec:Conclusions}Conclusions}
We have presented in this paper a low dimensional model for convection of an electrically conducting fluid enclosed between two rigid horizontal boundaries in a uniform vertical magnetic field. We find that the global quantities scale with the distance from criticality $\epsilon = r-1$ near the onset of primary instability. The kinetic energy $K$ , convective entropy $\Phi$ and convective heat flux $Nu-1$ scale with $\epsilon$ as $\epsilon$, $\epsilon^{0.9}$ and $\epsilon^{0.91}$ respectively. The vertical magnetic field inhibits the primary as well as secondary instabilities. Onset is always in the form of 2D stationary convection for the range of $Pr$ investigated. The secondary instability is time dependent, which may either be periodic or quasiperiodic. The model shows the appearance of standing waves at the onset of secondary instability. With further increase of $r$, there is a bifurcation to traveling wave solutions. A traveling wave solution consists of oblique wavy rolls whose axis alternate quasiperiodically in time between two directions perpendicular to each other. We find qualitative similarity of the results of this simple model with the results of DNS by Clever and Busse~\cite{cb89}. The rms value of the vertical velocity has one peak in the middle of the cell. The rms values for the horizontal velocities show bi-modal behaviour with two peaks are located symmetrically about the mid-plane ($z=0$). The rms value of the convective temperature is flat near the middle of the cell for stationary convection and become bi-modal for time dependent convection. The two peaks of the convective temperature are also located symmetric about the mid-plane. However, the locations of the peaks for the convective temperature is closer to the mid-plane compared with the same for the peaks of the horizontal velocities. The rms values of all fields vary with variation of $r$ and $Q$.

\section*{Acknowledgements}
We have benefited from fruitful discussions with Priyanka Maity and Hirdesh Pharasi.

%

\end{document}